\documentclass[aps,pra,twocolumn,superscriptaddress,nofootinbib]{revtex4-2}

\usepackage[utf8]{inputenc}
\usepackage[T1]{fontenc}
\usepackage{lmodern}
\usepackage{amsmath,amssymb}
\usepackage{graphicx}
\usepackage{booktabs}
\usepackage{pgfplots}
\pgfplotsset{compat=1.17}
\usepackage[colorlinks=true,linkcolor=blue,citecolor=blue,urlcolor=blue]{hyperref}

\begin{document}

\title{Class-resolved sifting for quantum key distribution with photon pairs hyperentangled in polarization and transverse spatial variables}

\author{Murilo H. Magiotto}
\affiliation{Instituto de F\'{\i}sica, Universidade Federal de Goi\'as, 74690-900, Goi\^ania, Goi\'as, Brazil}
\author{Millena B. Novaes}
\affiliation{Instituto de F\'{\i}sica, Universidade Federal de Goi\'as, 74690-900, Goi\^ania, Goi\'as, Brazil}
\author{Wesley B. Cardoso}
\affiliation{Instituto de F\'{\i}sica, Universidade Federal de Goi\'as, 74690-900, Goi\^ania, Goi\'as, Brazil}
\author{Ardiley T. Avelar}
\affiliation{Instituto de F\'{\i}sica, Universidade Federal de Goi\'as, 74690-900, Goi\^ania, Goi\'as, Brazil}
\author{Guilherme L. Zanin}
\affiliation{Instituto de F\'{\i}sica, Universidade Federal de Goi\'as, 74690-900, Goi\^ania, Goi\'as, Brazil}
\author{Rafael M. Gomes}
\email{rafaelgomes@ufg.br}
\affiliation{Instituto de F\'{\i}sica, Universidade Federal de Goi\'as, 74690-900, Goi\^ania, Goi\'as, Brazil}

\date{\today}

\begin{abstract}
Photon pairs hyperentangled in polarization and transverse spatial variables can carry two key bits per pair in entanglement-based quantum key distribution. We develop class-resolved sifting for such sources: every detected pair is assigned to a class according to which degrees of freedom were measured in compatible bases, and error and key rates are estimated separately in each class. Applied to previously reported data from a two-crystal spontaneous parametric down-conversion source read out in intermediate fractional-Fourier planes, the method recovers the key yield of independent sifting and turns the comparison between classes into a quantitative consistency test for cross-dependencies between the two degrees of freedom. The measured dependence lies below the spatial--polarization coupling predicted for the source, at the edge of the present statistical sensitivity. Resolving the data by polarization-analyzer setting, we find an approximately twofold dependence of the spatial error on the polarization setting in one readout arm, attributable to the measurement apparatus rather than to the source and relevant to basis-dependent side channels. We also compare the hyperentangled protocol with a single four-dimensional system measured in two mutually unbiased bases. Fundamentally, both yield the same key per detected pair and obey the same entropic bound, but they differ in their tensor-product structure: hyperentanglement factors the bases into separate two-outcome measurements, which are simpler to implement and allow errors to be attributed to each degree of freedom, whereas a single variable gives access to non-product bases that can increase the noise tolerance. For the present source, whose polarization coherence is limited by partial which-crystal erasure, a single-variable spatial encoding would yield more key only if its four-outcome readout kept the symbol error within a few percent.
\end{abstract}

\maketitle

%=====================================================================
\section{Introduction}
%=====================================================================

A photon pair hyperentangled in two degrees of freedom can feed two key channels at once, but only if each channel is secure in every round it is used. Hyperentanglement in polarization and transverse momentum is routinely produced by spontaneous parametric down-conversion (SPDC)~\cite{Kwiat1999,Barreiro2005}, and a recent experiment showed that a two-crystal source read out in intermediate fractional-Fourier planes keeps its polarization entanglement in every combination of spatial bases~\cite{Magiotto2026}. That work reported asymptotic rates of 0.56 and 0.84 secret bits per sifted pair in the polarization and spatial sectors, or about 1.4 bits when both bases match.

With independent basis choices in each degree of freedom, both bases match in only a quarter of the rounds, and in half of the rounds exactly one degree of freedom is measured compatibly. Treating each sector as an independent BBM92 channel~\cite{BBM92} uses these rounds implicitly. This is secure, within the effective-qubit measurement model, whether or not the errors of the two degrees of freedom are correlated, because the key rates of the two channels can be bounded jointly for any error statistics~\cite{Magiotto2026}. What independent sifting does not show is whether the errors of one degree of freedom depend on the basis chosen in the other. In the non-collinear two-crystal source such a dependence is physically plausible: the transverse variables carry which-crystal information, and the polarization coherence is zero in the crystal plane and partial in general intermediate planes~\cite{Magiotto2026}. In the intermediate planes used here the which-crystal information is only partially erased, and the residual coupling is predicted to reduce the polarization concurrence to about 0.93 when the spatial bases are compatible and 0.96 when they are not~\cite{Magiotto2026}.

Here we keep the rounds of each class apart and estimate their error rates separately. Our contributions are: (i) a class-resolved form of the sifting, in which every detected pair is assigned to one of four classes and contributes 2, 1, 1 or 0 key bits, with the error rates of each class estimated from that class alone; (ii) the yield per detected pair as a function of the basis bias, the gain of class-resolved over pooled estimation, and a joint four-dimensional bound for class A; and (iii) an evaluation on the previously reported data of Ref.~\cite{Magiotto2026}. That work introduced the underlying consistency test, which requires the polarization error to be the same in every combination of spatial bases and the spatial error to be the same for every setting of the polarization analyzers~\cite{Magiotto2026}. Here we embed this test in the sifting protocol, make it quantitative with exact tests and confidence intervals, compare it with the coupling predicted for the source, and resolve it by the setting of each polarization analyzer. Finally, (iv) we compare the hyperentangled encoding with a single four-dimensional variable, separate the fundamental from the technical differences, and determine the error rate below which each encoding yields more key for the present source.

%=====================================================================
\section{Class-resolved sifting protocol}
%=====================================================================

The protocol keeps the key bit of every degree of freedom measured in a compatible basis, and estimates the error rates of each class from that class alone. We label the polarization bases $Z_p=\{H,V\}$ and $X_p=\{D,A\}$. For the spatial sector, each party reads out the transverse variables in one of two fractional orders, $r$ or $s$, with a binary (two-slit) readout. Both readouts start from a fractional Fourier plane of order close to $45^\circ$; the $r$ readout images this plane onto the detector ($\alpha\simeq\pi/4$), and the $s$ readout places the detector in the focal plane of a further lens, which measures the transverse momentum at that plane ($\alpha\simeq3\pi/4$)~\cite{Magiotto2026}. The compatible configurations are those with opposite fractional orders. We therefore label Alice's readouts in orders $r$ and $s$ as $Z_s$ and $X_s$, and Bob's readouts in orders $s$ and $r$ as $Z_s$ and $X_s$, so that the compatible configurations $\{r^A,s^B\}$ and $\{s^A,r^B\}$ correspond to both parties choosing $Z_s$ or both choosing $X_s$. The same-order configurations $\{r^A,r^B\}$ and $\{s^A,s^B\}$ are incompatible.

Since the two orders differ by $\pi/2$, $r$ and $s$ are mutually unbiased in the same sense as position and momentum~\cite{Magiotto2026}. We treat the binary spatial readout as an effective qubit measurement with mutually unbiased bases, which is an idealization~\cite{Magiotto2026}. This description assumes that the two selected spatial cells define the relevant two-dimensional measurement subspace and that the associated postselection does not introduce an unaccounted basis-dependent detection bias, an assumption that we examine in Secs.~\ref{sec:results} and~\ref{sec:discussion}. Events outside the selected cells are discarded, so the resulting rates are conditional on the spatial filtering, and a security proof for the actual implementation must characterize the corresponding measurement operators and detection efficiencies. The protocol runs as follows: (1) the source sends one photon of each pair to Alice and one to Bob, and each party chooses a polarization basis and a spatial basis independently, taking the $Z$ basis with probability $p$ in each degree of freedom, and records both outcomes; (2) Alice and Bob announce their bases, and a degree of freedom is compatible when both parties chose $Z$ or both chose $X$; (3) each round is assigned to class A, B, C or D (Table~\ref{tab:classes}); (4) in each class, a random subset of the compatible outcomes in each basis is revealed to estimate the error rates $Q_z$ and $Q_x$ of the degree (or degrees) of freedom that carries key; and (5) error correction and privacy amplification are applied to the raw key of each class at the rate allowed by that class's own error statistics (Sec.~\ref{sec:rates}).

Outcomes of incompatible degrees of freedom are discarded because they are uncorrelated between distant parties, and since basis compatibility becomes public after step (2), neither carries a secret bit.

%=====================================================================
\section{Key rates and yield per detected pair}\label{sec:rates}
%=====================================================================

Each class is distilled at the rate set by its own error rates, and the yield per detected pair is the probability-weighted sum over classes. For a degree of freedom carrying key in class $k$, with error rates $Q_z^{(k)}$ and $Q_x^{(k)}$ in its two conjugate bases, we use the asymptotic lower bound for BBM92 with one-way error correction and privacy amplification~\cite{ShorPreskill2000,Gisin2002,Lo2005},
\begin{equation}
R^{(k)} \geq 1 - h\!\left(Q_z^{(k)}\right) - h\!\left(Q_x^{(k)}\right),
\label{eq:rate}
\end{equation}
where $h(x)=-x\log_2 x-(1-x)\log_2(1-x)$ is the binary entropy. This rate is conditional on the retained detection events and on the effective-qubit description of the spatial measurements, so the class-dependent rates are estimates within this measurement model rather than a finite-key security proof for the complete experimental apparatus. Class A carries one rate per degree of freedom, $R_p^{(A)}$ and $R_s^{(A)}$; classes B and C carry $R_p^{(B)}$ and $R_s^{(C)}$. With the same bias $p$ in both degrees of freedom, the key per detected pair is
\begin{equation}
Y(p) = q^2\left[R_p^{(A)}+R_s^{(A)}\right] + q(1-q)\left[R_p^{(B)}+R_s^{(C)}\right].
\label{eq:yield}
\end{equation}
Using class A alone gives $Y_A(p)=q^2[R_p^{(A)}+R_s^{(A)}]$. At $p=1/2$ the full protocol yields twice as much whenever the class B and C rates match those of class A. As $p\to1$ both expressions tend to $R_p^{(A)}+R_s^{(A)}$, but the $X$-basis rounds needed for parameter estimation become scarce, so the useful bias is set by a finite-key analysis~\cite{Lo2005,Tomamichel2012}.

\begin{table}[b]
\caption{Classes of the protocol. $q=p^2+(1-p)^2$ is the probability of compatible bases in one degree of freedom; $q=1/2$ for uniform choice.}
\label{tab:classes}
\begin{ruledtabular}
\begin{tabular}{ccccc}
Class & Polarization & Spatial & Prob. & Key bits\\
\hline
A & compatible & compatible & $q^2$ & 2\\
B & compatible & incompatible & $q(1-q)$ & 1 (pol.)\\
C & incompatible & compatible & $q(1-q)$ & 1 (spatial)\\
D & incompatible & incompatible & $(1-q)^2$ & 0\\
\end{tabular}
\end{ruledtabular}
\end{table}

In class A the two degrees of freedom can also be treated as a single four-dimensional system with the product bases $Z_p\otimes Z_s$ and $X_p\otimes X_s$, which are mutually unbiased. The entropic uncertainty relation with quantum memory~\cite{Berta2010} then gives
\begin{equation}
R^{(A)}_{\text{joint}} \geq 2 - H(E_{ZZ}) - H(E_{XX}),
\label{eq:joint}
\end{equation}
where $E_{ZZ}$ ($E_{XX}$) is the four-outcome error pattern (no error, polarization only, spatial only, both) of the rounds measured in $Z_p\otimes Z_s$ ($X_p\otimes X_s$) and $H$ its Shannon entropy. Because polarization and spatial bases are selected independently, class A contains four compatible product-basis configurations, which form two complementary pairs: $Z_p\otimes Z_s$ with $X_p\otimes X_s$, and $Z_p\otimes X_s$ with $X_p\otimes Z_s$. Bases within a pair are mutually unbiased, whereas bases from different pairs share a local basis and are not. Equation~\eqref{eq:joint} therefore applies to the first pair, and the second pair obeys the analogous bound with $E_{ZX}$ and $E_{XZ}$; each bound uses the error distribution estimated for its own pair, and the two contributions are weighted by the probabilities of their configurations within class A, $[p^4+(1-p)^4]/q^2$ and $2p^2(1-p)^2/q^2$. Since $H(E)$ is at most the sum of the marginal binary entropies, the bound for each pair, and hence their weighted combination, is never below $R_p^{(A)}+R_s^{(A)}$, and it gains when errors in the two degrees of freedom are correlated, as they are when a common background is present~\cite{Magiotto2026}.

Class resolution is not required for security: pooling the rounds of classes A and B (or A and C) and estimating a single pair of error rates gives a valid, slightly lower bound~\cite{Magiotto2026}. Because $h$ is concave, the class-weighted average of the rates of Eq.~\eqref{eq:rate} is never below the rate computed from the pooled error rates, with equality when the classes have the same error rates. The difference is of second order in the spread of error rates between classes and is quantified in Sec.~\ref{sec:results}. The main use of keeping the classes apart is that a difference between them reveals a dependence of one degree of freedom on the basis chosen in the other.

%=====================================================================
\section{Experimental setup}
%=====================================================================

\begin{figure}[t]
\centering
\includegraphics[width=\columnwidth]{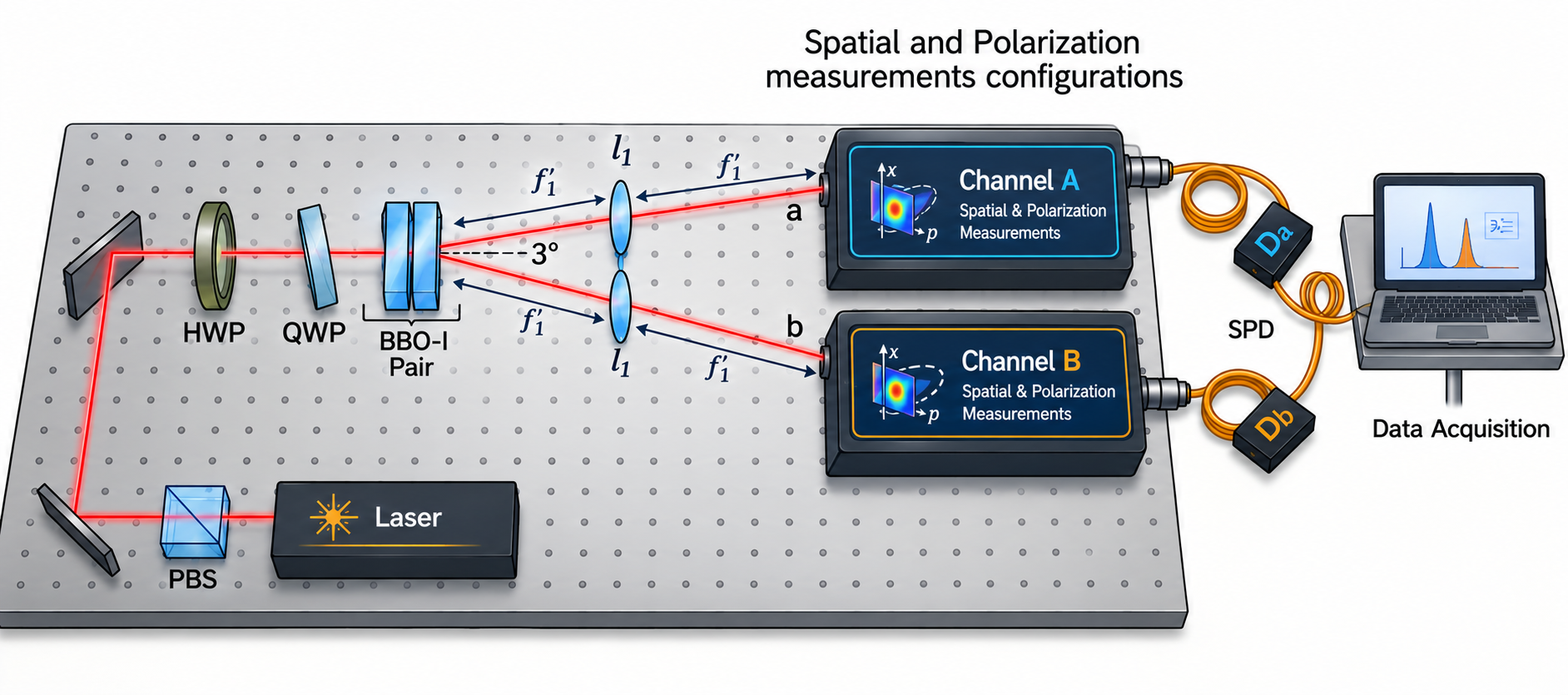}
\caption{Experimental setup of Ref.~\cite{Magiotto2026}. HWP: half-wave plate; QWP: quarter-wave plate; PBS: polarizing beam splitter; $l_1$: lens defining the fractional Fourier plane; $D_a$, $D_b$: single-photon detectors. Each channel contains the spatial readout optics and a polarization analyzer.}
\label{fig:setup}
\end{figure}

Figure~\ref{fig:setup} shows the experimental setup of Ref.~\cite{Magiotto2026}. A continuous-wave laser at 405~nm (50~mW), with its polarization set to diagonal by a polarizing beam splitter and a half-wave plate, pumps two adjacent 1-mm type-I BBO crystals with orthogonal optic axes, and a tilted quarter-wave plate compensates the relative phase between the two polarization components. The down-converted photons at 810~nm, emitted at about $3^\circ$ to the pump, pass through a lens $l_1$ ($f=1000$~mm) that defines, 300~mm beyond it, a fractional Fourier plane of order $45.6^\circ$. The $r$ readout images this plane onto the detector, and the $s$ readout places the detector in the focal plane of a further lens; slits of 0.60~mm ($r$) and 0.25~mm ($s$) select the spatial cells. Polarization is analyzed with a half-wave plate and a polarizing beam splitter in each arm, and the photons are coupled through 10-nm interference filters into multimode fibers and detected by one single-photon detector per arm, with a 2~ns coincidence window. The spatial cells are defined by scanning the detector of one arm while that of the other is held at two fixed positions. For each spatial configuration, all polarization settings were recorded sequentially within the same alignment, by rotating the half-wave plates. Changing the spatial configuration requires exchanging the second lens and realigning the system, so different spatial configurations were in some cases measured on different days; the acquisition time of each run was recorded, and differences between runs are larger than those within a run~\cite{Magiotto2026}.

%=====================================================================
\section{Results}\label{sec:results}
%=====================================================================

We apply the class-resolved analysis to the data of Ref.~\cite{Magiotto2026}, which cover all combinations of polarization and spatial bases. Table~\ref{tab:rates} lists the error rates and asymptotic key rates of each class. The following subsections examine these rates, the resulting yield per detected pair and the consistency between classes.

\subsection{Key rates per class}
In class A the joint bound of Eq.~\eqref{eq:joint} gives $1.430\pm0.027$ bits for the pair $(Z_p\otimes Z_s,\,X_p\otimes X_s)$ and $1.412\pm0.024$ for $(Z_p\otimes X_s,\,X_p\otimes Z_s)$. At uniform basis choice, where both pairs are equally likely, the weighted joint rate is $1.421\pm0.018$ bits, against $1.395\pm0.017$ for the sum of the separate rates (Table~\ref{tab:rates}). The gain is small because simultaneous errors in both degrees of freedom occur in only a few tenths of a percent of the rounds~\cite{Magiotto2026}. Within class B, the $\{r^A,r^B\}$ and $\{s^A,s^B\}$ configurations give $0.610\pm0.065$ and $0.609\pm0.037$ bits, which agree.

\begin{table}[b]
\caption{Error rates and asymptotic key rates per class, from the coincidence counts of Ref.~\cite{Magiotto2026}.}
\label{tab:rates}
\begin{ruledtabular}
\begin{tabular}{llcccc}
Class & DOF & $N_z/N_x$ & $Q_z$ (\%) & $Q_x$ (\%) & $R$ (bits)\\
\hline
A & pol.    & 8255/7939  & 1.89 & 5.49 & $0.558\pm0.013$\\
A & spatial & 10469/5725 & 1.14 & 0.89 & $0.837\pm0.011$\\
B & pol.    & 1380/1261  & 1.16 & 5.39 & $0.606\pm0.032$\\
C & spatial & 10254/5235 & 0.75 & 1.20 & $0.842\pm0.011$\\
\end{tabular}
\end{ruledtabular}
\end{table}

Pooling classes A and B in the polarization sector at $p=1/2$ gives 0.5813 bits per round, against 0.5819 for the class-weighted average of the separate rates. In the spatial sector, pooling classes A and C gives 0.8391 against 0.8395. Class-resolved estimation thus gains less than $10^{-3}$ bits here, because the error rates of the classes are close.

The polarization rate is limited by the conjugate-basis error, $Q_x=5.5\%$, which exceeds $Q_z$ by 3.6 percentage points. Since $Z_p$ outcomes do not depend on the coherence between the two crystal contributions, an excess in $X_p$ measures the incomplete polarization coherence, provided that background and the other measurement imperfections contribute approximately equally to both bases and that the relative phase between the two polarization components is compensated~\cite{Magiotto2026}. With a conjugate-basis visibility $(1-f)\,C$, where $f\simeq0.7\%$ is the background weight and $C$ the concurrence~\cite{Magiotto2026}, the concurrence $C\simeq0.93$ predicted for the compatible spatial configurations gives an excess of $(1-f)(1-C)/2\simeq3.5$ points, in agreement with the observed value. Under these assumptions the excess conjugate-basis error can be attributed predominantly to the partial erasure of the which-crystal information, and removing it would raise the polarization rate to about 0.73 bits per round.

\subsection{Yield per detected pair}

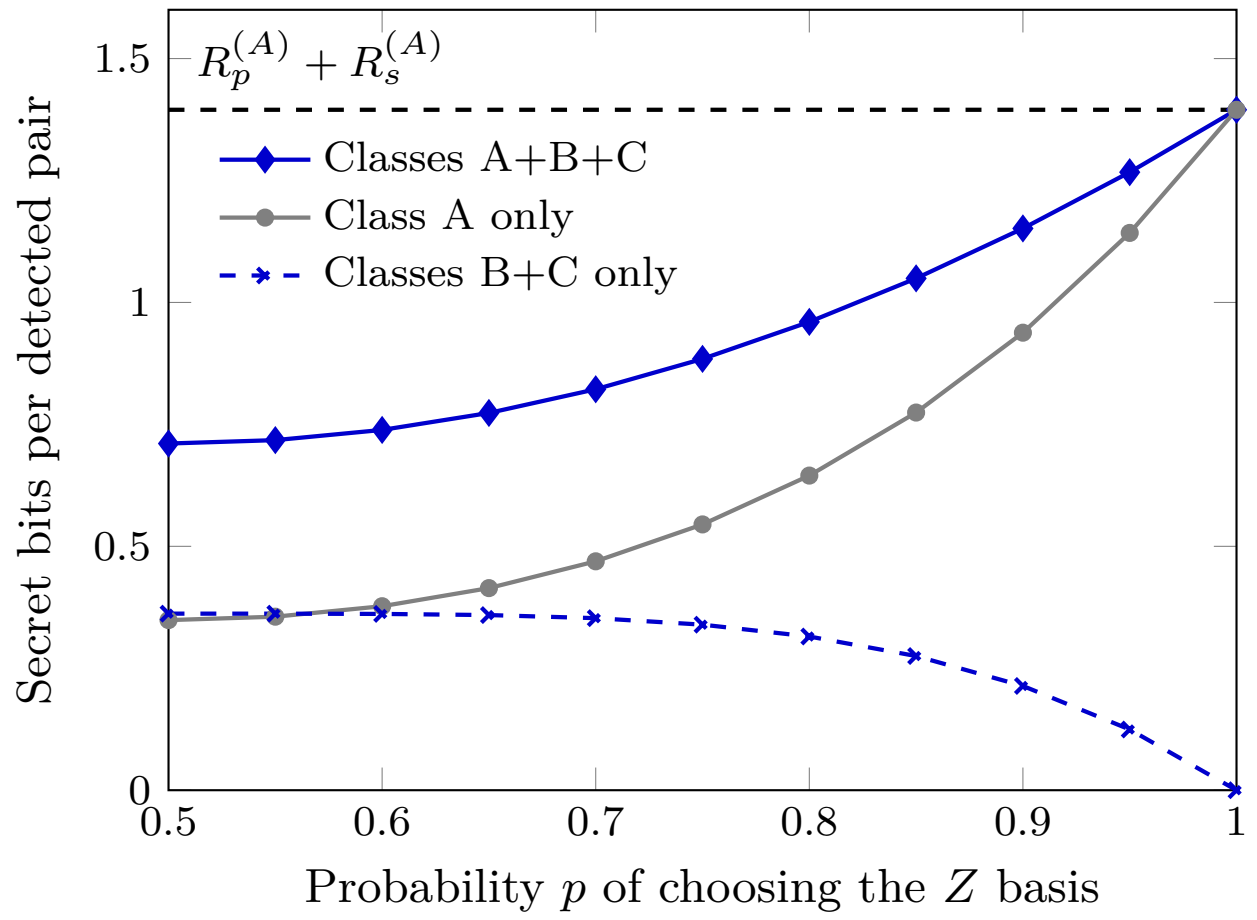
\begin{figure}[t]
\centering
\begin{tikzpicture}
\begin{axis}[
  width=\columnwidth, height=0.78\columnwidth,
  xmin=0.5, xmax=1, ymin=0, ymax=1.6,
  xlabel={Probability $p$ of choosing the $Z$ basis},
  ylabel={Secret bits per detected pair},
  xtick={0.5,0.6,0.7,0.8,0.9,1},
  ytick={0,0.5,1,1.5},
  legend style={at={(0.03,0.86)},anchor=north west,draw=none,fill=none,font=\footnotesize},
  legend cell align=left,
  tick label style={font=\footnotesize},
  label style={font=\small},
  declare function={
    qq(\p)=\p*\p+(1-\p)*(1-\p);
    RA=0.558+0.837;
    RBC=0.606+0.842;
  }
]
\addplot[blue!80!black, thick, mark=diamond*, mark size=2pt, samples=11, domain=0.5:1]
  {qq(x)^2*RA + qq(x)*(1-qq(x))*RBC};
\addlegendentry{Classes A+B+C}
\addplot[gray, thick, mark=*, mark size=1.3pt, samples=11, domain=0.5:1]
  {qq(x)^2*RA};
\addlegendentry{Class A only}
\addplot[blue!80!black, thick, dashed, mark=x, mark size=2pt, samples=11, domain=0.5:1]
  {qq(x)*(1-qq(x))*RBC};
\addlegendentry{Classes B+C only}
\addplot[black, thick, dashed, domain=0.5:1, samples=2, forget plot] {RA};
\node[anchor=south west,font=\footnotesize] at (axis cs:0.505,1.40) {$R_p^{(A)}+R_s^{(A)}$};
\end{axis}
\end{tikzpicture}
\caption{Asymptotic secret key per detected pair as a function of the probability $p$ of choosing the $Z$ basis in each degree of freedom, Eq.~\eqref{eq:yield}, with the class rates of Table~\ref{tab:rates}. Including classes B and C doubles the yield relative to class A alone at $p=1/2$ (0.71 vs 0.35 bits); the upper curve coincides with independent sifting in each degree of freedom. The dashed blue curve is the contribution of classes B and C alone, $q(1-q)[R_p^{(B)}+R_s^{(C)}]$, which equals that of class A at $p=1/2$. The dashed line is the class A rate per round, the limit of both curves as $p\to1$. Statistical uncertainties are at most 0.017 bits, the value reached at $p=1$.}
\label{fig:yield}
\end{figure}

Figure~\ref{fig:yield} shows Eq.~\eqref{eq:yield} evaluated with the class rates of Table~\ref{tab:rates}. Relative to keeping class A alone, the gain is largest at uniform choice and shrinks with the bias, since biased bases already make class A dominant. At uniform choice the full yield is $0.711\pm0.010$ bits per detected pair, twice the $0.349\pm0.004$ of class A alone; at $p=0.9$ it is 1.15 bits, 23\% more than class A alone (0.94). This gain is not an advantage over independent sifting, whose yield is the upper curve, since independent sifting already retains the compatible outcomes of each degree of freedom separately. The class-resolved formulation makes the contribution of the partially compatible rounds explicit and provides the diagnostic information discussed in Sec.~\ref{sec:consistency}.

Classes B and C together contribute $q(1-q)[R_p^{(B)}+R_s^{(C)}]$ bits per detected pair (dashed curve in Fig.~\ref{fig:yield}). At uniform choice this is $0.362\pm0.009$ bits, as much as class A; it falls to $0.214\pm0.005$ at $p=0.9$ and to zero as $p\to1$, where nearly every round falls in class A.

\subsection{Consistency between classes}
\label{sec:consistency}

\begin{table}[b]
\caption{Error rates of the same degree of freedom in different classes, their difference $\Delta$ (class A minus class B or C), and the two-sided $p$ value of Fisher's exact test. For reference, the four $p$ values correspond to 1.9, 0.1, 2.8 and $1.6\sigma$ of a two-sided normal test.}
\label{tab:consistency}
\begin{ruledtabular}
\begin{tabular}{lcccc}
Quantity & A (\%) & B or C (\%) & $\Delta$ (pp) & $p$\\
\hline
Pol.\ $Q_z$ (A vs B)     & 1.89 & 1.16 & $+0.73$ & 0.061\\
Pol.\ $Q_x$ (A vs B)     & 5.49 & 5.39 & $+0.10$ & 0.95\\
Spatial $Q_z$ (A vs C)   & 1.14 & 0.75 & $+0.39$ & 0.005\\
Spatial $Q_x$ (A vs C)   & 0.89 & 1.20 & $-0.31$ & 0.11\\
\end{tabular}
\end{ruledtabular}
\end{table}

We compare the error rates of two classes with Fisher's exact test~\cite{Agresti2013}, which gives the probability of observing a difference at least as large as the measured one if both classes shared the same error rate. Unlike the normal approximation behind the usual $z$ statistic, the test remains exact for small counts, which matters for class B, with only 16 computational-basis errors. It serves here as a diagnostic and plays no role in the security analysis, which uses only the error rates of each class.

Table~\ref{tab:consistency} compares the error rates of the same degree of freedom in different classes. The polarization error rates of classes A and B are statistically compatible in both bases ($p=0.06$ and $p=0.95$, Fisher exact test), and so are the conjugate-basis spatial error rates of classes A and C ($p=0.11$). The computational-basis spatial error rates of classes A and C differ ($p=0.005$). The polarization computational-basis difference has the same sign, with a larger error in class A, but is not significant with the statistics of class B ($p=0.06$); whether it shares the origin of the spatial difference cannot be decided from the present data. Reference~\cite{Magiotto2026} quotes the spatial error of class A only over the polarization outcomes without a polarization error (0.97\% and 0.72\%), which removes part of the background; here class A includes all polarization outcomes, as class C does, so that both carry the same background fraction.

The coupling predicted for the source affects only the conjugate polarization basis $X_p$, because $Z_p$ outcomes do not depend on the coherence between the two crystal contributions~\cite{Magiotto2026}. With the concurrences of about 0.93 for the compatible spatial configurations (class A) and 0.96 for the incompatible ones (class B), it predicts $Q_x^{(A)}-Q_x^{(B)}\simeq(1-f)(0.96-0.93)/2\simeq1.5$ percentage points (pp). The measured difference is $0.10$~pp, with a 95\% confidence interval of $[-1.4,+1.3]$~pp. We compute this interval with the Newcombe hybrid score method~\cite{Newcombe1998}, because class B contains few events, a regime in which the simple Wald interval is unreliable. The predicted difference thus lies just outside the interval, about $2.2\sigma$ above the measured value: the data do not show the predicted increase, but the present statistics are only just sufficient to probe it. The prediction is approximate, since it assumes collection apertures comparable to the size of the beams~\cite{Magiotto2026}, and the background-free ratios $V_{DA}/V_{HV}$ measured in Ref.~\cite{Magiotto2026} also show no systematic difference between compatible and incompatible spatial configurations, their variation being comparable to that between repeated alignments. In terms of the polarization visibility $V=1-2Q_x$, the data exclude a dependence on the spatial configuration larger than about 3 percentage points. The sensitivity is set by the 1261 $X_p$ events of class B; resolving a difference of 0.5~pp at the $2\sigma$ level would require about $1.6\times10^4$ $X_p$ events in each class. The reverse comparison, of the conjugate-basis spatial error in classes A and C, gives an interval of $[-0.71,+0.07]$~pp and excludes a dependence on the polarization configuration larger than about 0.7~pp. Since technical contributions add to any coupling in the source, these bounds also constrain the coupling itself, unless the two happen to cancel.

\begin{table}[b]
\caption{Spatial error in each spatial configuration as a function of the basis of the polarization analyzer in the arm read out in order $s$ and in the arm read out in order $r$, over all sixteen analyzer combinations recorded in Ref.~\cite{Magiotto2026}: errors/events, error rate in percent, and two-sided $p$ value of Fisher's exact test. Combined over the two configurations with a Mantel--Haenszel test, the $s$-arm effect has odds ratio 1.9 and $p=4\times10^{-8}$.}
\label{tab:settings}
\footnotesize
\begin{ruledtabular}
\begin{tabular}{llccc}
Config. & Arm & $Z_p$ & $X_p$ & $p$\\
\hline
$\{r^A,s^B\}$ & $s$ (Bob)   & 132/10418 (1.27) & 64/10305 (0.62) & $2\times10^{-6}$\\
              & $r$ (Alice) & 93/10270 (0.91)  & 103/10453 (0.99) & 0.57\\
$\{s^A,r^B\}$ & $s$ (Alice) & 76/5836 (1.30)   & 38/5124 (0.74)  & 0.004\\
              & $r$ (Bob)   & 53/5475 (0.97)   & 61/5485 (1.11)  & 0.51\\
\end{tabular}
\end{ruledtabular}
\end{table}

The computational-basis spatial difference arises between measurements taken within the same spatial alignment, so it is not caused by realignment. Reference~\cite{Magiotto2026} recorded the spatial error for all sixteen combinations of the polarization analyzers, which allows a finer analysis (Table~\ref{tab:settings}). The spatial error depends on the setting of the polarization analyzer in the arm read out in order $s$: in both spatial configurations it is about twice as large when that analyzer is set to $Z_p$ as when it is set to $X_p$ (Mantel--Haenszel odds ratio 1.9 across the two configurations, $p=4\times10^{-8}$~\cite{Agresti2013}), whereas the setting of the analyzer in the arm read out in order $r$ has no significant effect. This pattern is inconsistent with a source-only explanation: a $Z_p$ measurement by either party projects the pair onto the same crystal-selected spatial state, so a coupling in the source would not depend on which arm carries the $Z_p$ setting, whereas the observed effect follows the $s$ readout, which belongs to Bob in one configuration and to Alice in the other. This points to a setting-dependent effect associated with the measurement apparatus, rather than being explained by the source model alone, with implications for basis-dependent side channels. Temporal drift within a block cannot be completely excluded with the available acquisition record, although it would mimic the pattern only if the settings of the $s$-arm analyzer had been grouped in time in both configurations. When classes A and C are compared at the same $s$-arm setting, the difference is significant in one configuration (Mantel--Haenszel odds ratio 1.5, $p=0.004$)~\cite{Agresti2013} and of opposite sign in the other (0.74, $p=0.13$), with no consistent effect overall (1.17, $p=0.20$). The difference between the computational-basis spatial error rates of classes A and C, aggregated over analyzer settings, therefore cannot be attributed to an intrinsic class dependence: it is sensitive to the measurement configuration and should not be interpreted as evidence of a source-induced spatial--polarization coupling. A dependence of the error of one degree of freedom on the basis setting of the other is, however, precisely the kind of basis-dependent behavior that must be monitored, and bounded, in key distribution~\cite{Makarov2006,Fung2009}.

%=====================================================================
\section{Hyperentangled versus single-variable encoding}\label{sec:comparison}
%=====================================================================

The hyperentangled protocol uses a four-dimensional space per photon built from two qubits. The natural alternative is a single degree of freedom of the same dimension, for instance the transverse spatial variables read out in four bins per fractional plane and measured in two mutually unbiased bases. Table~\ref{tab:comparison} summarizes the comparison.

\begin{table}[b]
\caption{Comparison between the hyperentangled encoding (polarization and a binary spatial readout) and a single spatial variable read out in four bins, both measured in two mutually unbiased bases.}
\label{tab:comparison}
\footnotesize
\begin{ruledtabular}
\begin{tabular}{p{0.34\columnwidth}p{0.27\columnwidth}p{0.27\columnwidth}}
 & Hyperentangled & Single variable\tabularnewline
\hline
\raggedright Key per detected pair & \raggedright $2q$ & \raggedright $2q$\tabularnewline
\raggedright Entropic bound & \raggedright Eq.~\eqref{eq:joint} & \raggedright Eq.~\eqref{eq:joint}\tabularnewline
\raggedright Unbiased bases with local two-outcome readouts & \raggedright at most 3 & \raggedright none; up to 5 with four-outcome readouts\tabularnewline
\raggedright Spatial bins per plane & \raggedright 2, wide & \raggedright 4, narrower\tabularnewline
\raggedright Partially compatible rounds & \raggedright carry key (classes B, C) & \raggedright do not exist\tabularnewline
\raggedright Errors attributable to each degree of freedom & \raggedright yes & \raggedright no\tabularnewline
\raggedright Main limitation in the present source & \raggedright polarization coherence & \raggedright errors and geometry of narrow bins\tabularnewline
\end{tabular}
\end{ruledtabular}
\end{table}

With two bases chosen with the same bias, the bases of the two parties in the single-variable encoding coincide with probability $q$, and each coincident round carries $\log_2 4=2$ bits, so the yield is $2q$ bits per detected pair, as in Eq.~\eqref{eq:yield} with equal rates. Both encodings obey the same entropic bound, Eq.~\eqref{eq:joint}~\cite{Berta2010,Cerf2002}. Under isotropic noise, which spreads errors uniformly over the three wrong outcomes, both tolerate a symbol error of about 18.9\% when this joint bound is used, whereas bounding the two degrees of freedom separately lowers the threshold of the hyperentangled encoding to about 16.5\%. The two encodings differ in their tensor-product structure. In the hyperentangled system, the bases measured with separate local two-outcome readouts are tensor products of local qubit bases, and at most three of them can be mutually unbiased, since each qubit has at most three mutually unbiased bases; the remaining bases of a complete set of five require joint measurements of polarization and spatial variables. A single variable gives direct access, in principle, to the complete set, which increases the noise tolerance at the cost of a lower sifting probability~\cite{Cerf2002}. Conversely, only the hyperentangled system offers a physical partition into subsystems, so that errors can be attributed to each degree of freedom, as in the diagnostic of Sec.~\ref{sec:results}, and partially compatible rounds still carry key.

The hyperentangled readout separates the four-dimensional measurement into two two-outcome measurements: polarization, analyzed with a wave plate and a polarizing beam splitter, and the spatial variables, resolved into two wide bins. This tensor-product structure allows the key contribution of each degree of freedom to be estimated separately, so a reduction in the key rate of one degree of freedom does not necessarily eliminate the contribution of the other. This advantage does not require the two measurement channels to be operationally independent; possible cross-dependencies must instead be characterized and bounded, as illustrated by the analyzer-dependent spatial errors of Sec.~\ref{sec:consistency} (Table~\ref{tab:settings}). A four-bin spatial readout of the same transverse field instead requires narrower bins, which enlarge the errors caused by the finite width of the spatial correlations, and keeping its two readouts mutually unbiased, which for coarse-grained continuous variables requires a periodic bin geometry~\cite{Tasca2018}, constrains the alignment more tightly than in the binary case. Its advantages are a single degree of freedom to align and stabilize, a dimension that grows with the number of bins, and the absence of the coupling between degrees of freedom that limits the polarization coherence of the present source. Polarization, in turn, doubles any spatial dimension $d_s$ to $2d_s$ with standard optics, so the two approaches can also be combined.

Which encoding yields more key depends on the error rates that each achieves. In the present source, class A yields $R_p^{(A)}+R_s^{(A)}\simeq1.40$ bits per compatible round with separate bounds and 1.42 with the joint bound, limited mainly by the conjugate-basis polarization error that the partial erasure of which-crystal information leaves (Sec.~\ref{sec:results}). A spatial ququart with the same symbol error $e$ in both bases, spread uniformly over $k$ of the three wrong outcomes, yields $2-2[h(e)+e\log_2 k]$ bits per compatible round. It would provide a higher asymptotic key rate per compatible round only if $e$ stayed below a break-even value between about 3.7\% and 5.4\%, depending on $k$ and on whether the separate or the joint bound is used for the hyperentangled rate. For comparison, the binary spatial readout of this source has error rates of about 1\%. Whether a four-bin readout can stay below the break-even value depends on the width of the spatial correlations and on the bin geometry, and remains to be established experimentally; its detection efficiency would in addition set the key per emitted pair. The hyperentangled encoding becomes preferable for sources whose polarization coherence is closer to ideal, and in any case it retains the robustness and diagnostic advantages described above.

%=====================================================================
\section{Discussion}\label{sec:discussion}
%=====================================================================

Class-resolved sifting does not increase the key yield beyond that obtained by independently sifting each degree of freedom. Its main value lies in the additional physical diagnostics provided by the class-dependent statistics and in a slightly tighter security accounting. Rounds in which only one degree of freedom has compatible bases can also contribute to the key, yielding 0.71 bits per detected pair under uniform basis choice and 1.15 bits for $p=0.9$, compared with 0.35 and 0.94 bits, respectively, when only class A is retained. Independent sifting achieves the same yields.

Beyond key generation, resolving the data by class provides a built-in diagnostic of spatial--polarization coupling during key distribution. Comparing the error rates of the same degree of freedom across classes formalizes the first two consistency conditions of Ref.~\cite{Magiotto2026}, which the joint error statistics of that work complement, and turns them into a continuous consistency check during key distribution. With the present data the check excludes a dependence of the conjugate-basis polarization error on the spatial configuration larger than about 1.4 percentage points, and of the conjugate-basis spatial error on the polarization configuration larger than about 0.7 points. The coupling predicted for the source lies at the edge of this sensitivity, which improves with the square root of the number of class-B and class-C events. Asymptotically, class-resolved estimation can only improve the rate through the concavity of the binary entropy $h$, and the gain is below $10^{-3}$ bits per round in the present case. At finite block sizes, splitting the parameter-estimation data among classes enlarges the statistical fluctuation of each estimate, which is likely to outweigh this gain unless the classes exhibit substantially different error rates. The diagnostic is therefore best applied on top of pooled parameter estimation rather than as a replacement for it.

A complementary approach is to treat the two degrees of freedom jointly as a single four-dimensional system. For class A, this joint bound increases the asymptotic key rate by about 0.03 bits per round. However, for the dominant pair $(Z_p\otimes Z_s,\,X_p\otimes X_s)$ its parameter estimation requires rounds measured in $X$ in both degrees of freedom, which are only a fraction $(1-p)^2$ of the polarization test rounds, 1\% at $p=0.9$. At the biased basis choices favored at finite block sizes, the statistical cost of this small sample is likely to outweigh the gain. More generally, for $d$ degrees of freedom, the classes correspond to the $2^d$ possible patterns of basis compatibility. The same accounting framework therefore extends to hyperentangled sources involving polarization, path, orbital angular momentum, or time-bin degrees of freedom, as well as to implementations using a larger number of transverse spatial cells per photon.

The present analysis has certain limitations that we intend to address in future work. (i) The class-B polarization comparison rests on 1261 conjugate-basis events, distributed between the two incompatible spatial configurations, with fewer than 400 in the $\{r^A,r^B\}$ configuration. An acquisition with about an order of magnitude more class-B events, which would bring its uncertainty to the level of class A, is therefore needed to decide whether the predicted coupling is present. (ii) The spatial configurations were measured in separate runs, after realignment and in some cases on different days, so the yields of Eq.~\eqref{eq:yield} are projections for a protocol with random basis choice rather than the output of such a protocol; the data were also recorded with scanning single detectors, as a characterization of the source~\cite{Magiotto2026}. (iii) A finite-key analysis, which would fix the optimal basis bias and the share of the key supplied by classes B and C at a given block size, is not attempted here; it must account for the fluctuation of all estimated error rates and for the abort probability, and requires a bound valid in the entanglement-based setting~\cite{Tomamichel2012}. (iv) The two spatial bases are read out with different optics and slit widths, so their detection efficiencies need not be equal~\cite{Magiotto2026}, and the dependence of the spatial error on the analyzer setting in the $s$ arm (Table~\ref{tab:settings}) shows that the readout of one degree of freedom is not fully independent of the basis setting of the other. Such setting-dependent measurement behavior represents a potential side-channel vulnerability in a practical implementation~\cite{Makarov2006,Fung2009}. Its security implications cannot be quantified from the error-rate differences alone and require a characterization of the setting-dependent detection efficiencies and measurement operators. In the class-A data, the ratio between the coincidence counts of the two spatial cells changes by about 10\% with the $s$-arm setting in the $\{r^A,s^B\}$ configuration and not at all in the $\{s^A,r^B\}$ one, although these counts, recorded at different times, are also affected by drift of the pump power. With that characterization, a security analysis with characterized efficiency mismatch~\cite{Fung2009,Zhang2021}, or a numerical key-rate calculation with the measured measurement operators~\cite{Winick2018}, would turn this observation into a bound on the key rate; it requires the coincidence rates normalized to the acquisition time and the pump power, and a direct measurement of the beam position at the $s$-readout slit for each analyzer setting.

%=====================================================================
\section{Conclusion}
%=====================================================================

Class-resolved sifting sorts the rounds of a hyperentangled key distribution by which degrees of freedom were measured compatibly and estimates the errors of each class separately. Applied to the previously reported data of a two-crystal SPDC source read out in intermediate fractional-Fourier planes~\cite{Magiotto2026}, it gives 0.71 secret bits per detected pair at uniform basis choice and 1.15 at a bias of 0.9: the same yield as independent sifting, and twice what rounds with both bases compatible provide at uniform choice.

Class resolution does not raise the key yield, but it provides a quantitative diagnostic of possible spatial--polarization dependencies. Within the available statistics, the conjugate-basis polarization error shows no significant dependence on the spatial configuration, whereas the source model of Ref.~\cite{Magiotto2026} predicts a difference of about 1.5 percentage points, just outside the 95\% confidence interval of the measurement. The present data therefore constrain, but do not rule out, the predicted coupling, and an acquisition with about an order of magnitude more class-B events would decide it. Conversely, the analyzer-resolved analysis reveals an approximately twofold dependence of the spatial error on the polarization setting in the $s$-arm readout, which points to a measurement-system contribution that must be characterized and bounded in a practical QKD implementation.

The comparison with a single four-dimensional variable shows that the two encodings are equivalent in key per detected pair and in their entropic bound, and differ in tensor-product structure. For the present source, a spatial ququart would yield more key only if its four-bin readout kept the symbol error below about 3.7--5.4\%, whereas hyperentanglement offers simpler two-outcome readouts, a separate key contribution from each degree of freedom and the class-resolved diagnostic, and becomes preferable as the polarization coherence of the source approaches the ideal.

\begin{acknowledgments}
This work was supported by the Brazilian funding agencies: Conselho Nacional de Desenvolvimento Cient\'{\i}fico e Tecnol\'ogico (CNPq) [DOI 10.13039/501100003593, Grant No.~306105/2022-5, Grant No.~312173/2022-9, Grant No.~303145/2026-9], Coordena\c{c}\~ao de Aperfei\c{c}oamento de Pessoal de N\'{\i}vel Superior (CAPES) [DOI 10.13039/501100002322], Funda\c{c}\~ao de Amparo \`a Pesquisa do Estado de Goi\'as (FAPEG) [Grant No.~LAM2026011000041, Grant No.~202510267001843], Instituto Nacional de Ci\^encia e Tecnologia de Infraestruturas Qu\^antica e Nano para Aplica\c{c}\~oes Convergentes (406636/2022-2), and Instituto Nacional de Ci\^encia e Tecnologia de Dispositivos Qu\^anticos (408783/2024-9).
\end{acknowledgments}

\end{document}